# GOEDEL'S OTHER LEGACY AND
## THE IMPERATIVE OF A SELF-REFLECTIVE SCIENCE

V. Basios (*),  E. Bouratinos (**)

(*)Vasileios Basios  is a physicist working as a researcher in physical chemistry, nonlinear science and complex systems at the University of Brussels (ULB). Contact: *vbasios@ulb.ac.be*
(*) Emilios Bouratinos is an essayist and writer on philosophy. He lives and works in Athens and Oxford. Contact: *ebouratinos@hol.gr*

ABSTRACT
The Goedelian approach is discussed as a prime example of a science towards the origins.  While mere self-referential objectification locks in to its own by-products, self-releasing objectification informs the formation of objects at hand and their different levels of interconnection. Guided by the spirit  of Goedel's work a self-reflective science can open the road where old tenets see only blocked paths.

*"This is, as it were, an analysis of the analysis itself, but if that is done it forms the fundamental of human science, as far as this kind of things is concerned."*  G. Leibniz, ('Methodus Nova ...', 1673)

### I. Introduction

It is reported that the great mathematician Kleene used to ask his doctorate candidates to "name five theorems of Goedel". Professor Kleene's point was that every and each one of these theorems had opened a whole new branch of mathematics and/or mathematical logic, all indispensable  for  the education of the new generation. The 'protean character' of the incompleteness theorems of him render themselves to an ongoing reinterpretation within purely mathematical but also in quite practical computing science context[1]. Goedel's legacy [1, 2] only lately widely recognized after a series of path-breaking studies, was destined to enter to an array of diverse fields such as consciousness studies [3], algorithmic complexity [4], mathematical physics -especially with the questions on the nature of space and time,  to mention a few, and even epistemology and philosophy proper. The spirit and vision of Goedel's work merits the broadest appreciation possible.

To the student of Goedel's methods and thinking at large it comes without surprise how his vision informed his virtuosity. It suffices to recall that in the heart of his epochal paper of 1931 lies his key idea to modify the ancient "antinomy of the liar" coming to us via Eubulides in the form of the affirmative "This sentence is false".  Through this gender of paradoxes, from the 'sorites paradoxes' of Zeno of Elea and his teacher Parmenides, to Socrate's method of 'aporia', one traces the perennial quest of illuminating the relation between truth and provability, or between reality ('aletheia') and our understanding of truth ('doxa', paradox itself literally meaning against 'doxa' or dogma). Science stands as the prime example of this continuous struggle between perceptions of reality and reality itself.

### II. Symbols, Objects, Language, Meaning

*"the certainty of mathematics is to be secured not by ... the manipulation of physical symbols, but rather by cultivating ... knowledge of the abstract concepts themselves"* [what is needed is]

---

1  The diagonalization argument has been reapplied for example in the demonstration of the Paris-Harrington theorem (*Handbook of Mathematical Logic*, pp. 1133-42) as early as 1977  and since then to other proofs of undecidability but also in virus detection program design *Am. Math. Monthly* **96**, pp. 835-6 (1989).

*"a clarification of meaning that does not consists of giving definitions"* K. Goedel [3], pg219

Goedels' thinking demands to complement our study of nature with the nature of this study. It also demands to maintain alive a thread of meaning through abstractions, mental objectifications and emerging ultrastructures and keep the awareness of where we reduce from that which we reduce to.

Ever since the Enlightment, there has been a conviction that *reason* equates *logos*. It goes without saying ofcourse that the original usage of logos appears to have been richer than the current usage of rational understanding. The moment has come to revise this opinion. New thinking, particularly in theoretical physics, epistemology, (meta)mathematics, and neurophysiology of the brain mandate it. Denying the equation *logos-reason* in the name of current scientific thinking doesn't imply a rejection of *logos* or *reason*. It implies a return to the original usage of *logos*, which was to penetrate the image, the processes and the blank spots of reality[2]. It also implies a return to the original scientific spirit utilizing the facts in going beyond the facts. As a colleague once put it 'the bones of dinosaurs tell us about the dinosaurs not about the bones of dinosaurs'.

Or to put it in a more poetic way: "One has to transcend the limitations of language through language itself, in order to express the higher layers of truth"[3]. To which we would add that unless one *is* able to find fitting ways for "expressing the higher layers of truth", he cannot even describe the lower levels of reality. The sensitivity required for doing the first is the same as that required for doing the second.

## II. Self-locking & Self-releasing objectification

*"I know one thing that I know no thing."* Socrates

Language does, of course, point to specific objects. It does so, however, only to the extent that the individual apprehended the specific object as informed by the whole, which merely expressed itself through the words. To comprehend the new and at the same time ancient question of understanding nature and the nature of this understanding it is necessary to explore in depth the process employed by the latter when it engages in objectification. The ability of our mind or consciousness in curving out and abstracting objects as separate from their surrounding whole, what we call objectification, has the natural tendency to 'self-lock' in its creations. Among a plethora of such instances the process of mechanization[4] of thought and nature is one of them.

At its high point in cartesian/newtonian thinking mechanized objectification has chased away lived experience, and we are all aware of its cruel outcome for nature, animals (human and non-human alike) and societal issues. Values and meaning expelled to the realm of unnecessary metaphysics. Today this attitude is less prevalent. We owe to Kurt Goedel, who was deeply preoccupied with the ultimate questions of understanding, that he showed to us that it is impossible to evict value judgement from even mathematical logic the most abstract of all disciplines. We need to be guided by both new research and new thinking loyal to the tradition informing Goedel and many other modern thinkers. Another kind of objectification, that of what we call "self-releasing" is at work here. Here, critical thinking, self-reflection and keeping in mind the different levels present in a complex interrelated whole illuminates both objects and their interrelationships. Only by being aware of the creative interplay between different levels of descriptions illuminates how one emerges from the other and how they complement each other.

---

2 Ancient Greek is shot through with such a fluid understanding of reality. 'Quality' (poiotis) came from 'doing' (poiein), 'thing' (pragma) came from 'acting' (prattein), 'truth' (aletheia) came from 'not-slumbering' (a-lethe) and 'consciousness' (syneidesis) came either from 'bringing together two knowable forms' (syn-eidos), or from adding something indefinable to information. Heraclitus, echoing the old function of language, says that Apollo "neither reveals nor conceals. He gives a sign."

3 We owe this expression to our Indian friend, scholar and poet Prof. Vishnu Narayan Namboodiri

4 Interestingly enough "mechane" a Homeric-greek word, where "machine" is coming from, originally meant a device of the mind to resolve problems (as in "deus ex machina"). The word "mechanized', itself a 'victim' of self-locking obectification, came to mean a mindless automated process!

No wonder that objecthood no longer appears as an attribute of things, but only of the way in which things are apprehended. From their different perspectives quantum theory, relativity and more recently nonlinear science and complex system studies have all dereifed the world. As a result we are now able to view the latter as an ongoing fusion[5] of physical entities among themselves on the one hand—and of observers with physical entities on the other. Physics in particular has lost a good deal of its nineteenth century arrogance. It no longer permits the cartesian/newtonian approach to dictate all of its thinking, by the sustained use of critical intellect, today physics boldly goes beyond physics and even managed to push itself to the forefront of the consciousness debate.

### III. Understanding beyond Paradigms

*" … mind, in its use, is not static, but constantly developing."*
*"Life force is a primitive element of the universe and it obeys certain laws of action.*
*These laws are not simple and not mechanical."* K. Goedel, [1], pp 233 & 240.

If 20[th] century physics have taught us one thing, it is that its findings point way beyond its conceptual framework. It doesn't mean we need a new epistemology or a new paradigm. What we need is a pre-epistemology and a non-paradigm. We must learn to think in terms of what we apprehend. We must stop apprehending in terms of what we think.

The basic task is to become aware of how we objectify the world -- and why we lock into our objectifications once we have done so. Pre-epistemology will help us understand nature in more subtle ways. A non-paradigm will help us avoid getting stuck on any conception of it. Quality cannot be appreciated by a person obsessed with quantity; non-local connections don't reveal themselves to localising mindsets; dynamic processes are not accessible to a structure-mediated worldview.

Nothing of this means we should discard the tools of modern science. It only means their use should become more discerning. Ultimately three things matter: (1) that we keep systems and minds open; (2) that in fragmenting and abstracting nature we never lose sight of its oneness; (3) that what we count doesn't dictate for us what counts.

The first task of a self-reflecting science is to define its specific goals, qualities and interaction with the rest of science and the world in *open-ended* terms. In 1972, Goedel raised a question that for him seem to bear most critical answers for the future development of science. He asked whether our physical and biochemical substratum permits a mechanical interpretation of all the functions of life and the mind. It touched upon on the nature of consciousness, life and mind in a very concrete way. Indeed any possible answer to this question hinges upon how complexity emerges in quantum systems, the borderline between quantum theory and biochemistry and the application of algorithmic complexity theory to quantum information theory. What is at stake here requires an interdisciplinary effort of immense proportions.

How do we understand life, mind and whether these are cosmic phenomena or mere earthbound accidents depend on the outcome of this question. But principally their outcome depends upon how we pose these questions. The only way to seek an understanding is being aware of our biases, free of our preconceptions undistracted of epistemological dogmas or mental blocks and paradigmatic thinking and able to reflect upon our own understanding. If consciousness, life and mind are primordial elements of the universe as Kurt Goedel maintained then we might be able to let them free to reflect upon themselves through our science.

---

5  The term 'fusion' hints a little more appropriately at the actual state of reality described as 'fuzzy' 'A' fuses into 'Non-A' and vice versa. 'A' also fuses (however slightly) with 'B', 'C' and all other entities. 'Fuzziness' merely describes passively the indistinctiveness of the borderline among entities and between entities and non-entities. 'Fusion" describes actively how each component constantly interpenetrates with the others.